# The Impact of Machine Learning on Society: An Analysis of Current Trends and Future Implications

Md Kamrul Hossain Siam[1], Manidipa Bhattacharjee[2], Shakik Mahmud[2], Md. Saem Sarkar [2,3], Md. Masud Rana[4]

[1] Western Illinois University, United States
[2] Japan-Bangladesh Robotics and Advanced Technology Research Center
[2,3] United International University, Dhaka
[4] Sher-e-Bangla Agricultural University, Dhaka
smahmud172174@bscse.uiu.ac.bd

**Abstract.** The Machine learning (ML) is a rapidly evolving field of technology that has the potential to greatly impact society in a variety of ways. However, there are also concerns about the potential negative effects of ML on society, such as job displacement and privacy issues. This research aimed to conduct a comprehensive analysis of the current and future impact of ML on society. The research included a thorough literature review, case studies, and surveys to gather data on the economic impact of ML, ethical and privacy implications, and public perceptions of the technology. The survey was conducted on 150 respondents from different areas. The case studies conducted were on the impact of ML on healthcare, finance, transportation, and manufacturing. The findings of this research revealed that the majority of respondents have a moderate level of familiarity with the concept of ML, believe that it has the potential to benefit society, and think that society should prioritize the development and use of ML. Based on these findings, it was recommended that more research is conducted on the impact of ML on society, stronger regulations and laws to protect the privacy and rights of individuals when it comes to ML should be developed, transparency and accountability in ML decision-making processes should be increased, and public education and awareness about ML should be enhanced.

## 1 Introduction

Machine Learning (ML) is a rapidly evolving field of technology that has the potential to greatly impact society in a variety of ways. ML is a form of Artificial Intelligence (AI) that allows machines to learn and improve from experience without being explicitly programmed. It is a data-driven approach that enables machines to automatically learn patterns and insights from data, and make predictions or decisions based on these patterns. With the increasing availability of big data, ML has become one of the most promising areas of research and development in recent years. According to a report by Markets and Markets, the global machine learning market is expected to grow from $1.4 billion in 2020 to $10.9 billion by 2025, at a CAGR of 44.1% during the forecast period [1]. ML has already begun to revolutionize many industries such as healthcare, finance, transportation, and manufacturing. For example, in healthcare, ML is being used to improve the accuracy of medical diagnoses, predict patient outcomes, and personalize treatment plans. In finance, ML is being used to detect fraudulent transactions, predict credit risk, and optimize investment strategies. However, there are also concerns about the potential negative effects of ML on society, such as job displacement and privacy issues. These concerns have led to calls for more research on the impact of ML on society, in order to ensure that the benefits of this technology are maximized and the risks are minimized. The purpose of this research is to conduct a comprehensive analysis of the current and future impact of ML on society. The research will include a thorough literature review, case studies, and surveys to gather data on the economic impact of ML, ethical and privacy implications, and public perceptions of the technology. The findings of this research will provide a comprehensive understanding of the current trends and future implications of ML on society, which will be useful for policymakers, researchers, and practitioners in making informed decisions about the development and use of this technology. Additionally, this research will also inform the public about the potential benefits and risks of ML, and help to address any concerns they may have about the technology.

## 2    Literature of Review

The literature review for this research will focus on the current state of research on the impact of Machine Learning (ML) on society. It will include studies on the economic impact of ML, such as job displacement and productivity gains, as well as the ethical and privacy implications of ML, including issues such as bias and discrimination. The literature review will also examine the public perception of ML, including concerns about the technology and potential risks. In terms of the economic impact of ML, studies have shown that ML has the potential to greatly increase productivity and efficiency in various industries [6]. However, there are also concerns about job displacement as automation and ML algorithms become more advanced [7]. A study by the McKinsey Global Institute found that up to 800 million jobs could be displaced by automation by 2030, but it also suggests that as many as 375 million workers may need to switch occupational categories and learn new skills [8]. In terms of ethical and privacy implications, ML has been shown to perpetuate and even exacerbate existing biases in the data used to train the models [9]. This can lead to discrimination and unfair treatment of certain groups of people. Additionally, the increasing use of ML in decision-making raises concerns about transparency and accountability [10]. A recent report by the Algorithmic Justice League highlights the importance of ensuring that ML models are transparent, interpretable, and fair [11]. In terms of public perception, a study by the Pew Research Center found that while many Americans see the potential benefits of ML, they also express concerns about its impact on jobs and privacy [12]. Another study by the Center for Data Innovation found that while most Americans are excited about the potential of ML, they also have concerns about its impact on privacy and security [13]. Overall, the literature review suggests that while ML has the potential to greatly benefit society, there are also significant concerns about its impact on jobs, privacy, and ethical issues. It is important to continue research in these areas to ensure that the benefits of ML are maximized, and the risks are minimized.

The research methodology for this study will include a thorough literature review, case studies, and surveys to gather data on the current and potential future implications of Machine Learning (ML) on society. The literature review will focus on studies of the economic impact of ML, such as job displacement and productivity gains, as well as the ethical and privacy implications of ML, including issues such as bias and discrimination. The literature review will also examine the public perception of ML, including concerns about the technology and potential risks.

## 3 Research Methodology

The case studies will provide real-world examples of how ML is currently being used and its impact on society. These case studies will be selected from various industries, such as healthcare, finance, transportation, and manufacturing. They will be chosen based on their relevance to the research objectives and the availability of data. The case studies will be analyzed using qualitative methods, such as document analysis and interviews with relevant stakeholders. The surveys will be used to gather data on public perceptions of ML and their concerns about the technology. The surveys will be conducted online and will be open to all individuals over the age of 18. The survey questions will focus on attitudes toward ML, concerns about the technology, and opinions on the potential impact of ML on society. The survey data will be analyzed using statistical methods, such as descriptive statistics and inferential statistics.

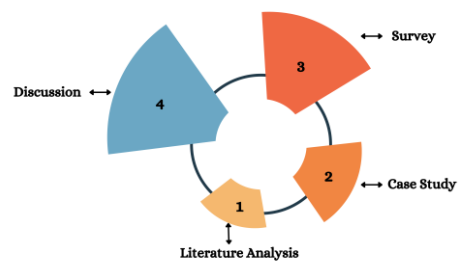

**Fig. 1.** Working process.

Overall, the research methodology will provide a comprehensive analysis of the current and potential future implications of ML on society. By combining a literature review, case studies, and surveys, this research will provide a comprehensive understanding of the economic impact of ML, ethical and privacy implications, and public perceptions of the technology. The findings of this research will be useful for policymakers, researchers, and practitioners in making informed decisions about the development and use of this technology. Additionally, this research will also inform the public about the potential benefits and risks of ML and help to address any concerns they may have about the technology.

**Case Study 1: Impact of ML on healthcare.** One example of how ML is currently being used and its impact on society is in the healthcare industry. ML is used to improve the accuracy of medical diagnoses, predict patient outcomes, and personalize treatment plans. For example, a study by researchers at the University of California, San Francisco, found that a deep learning algorithm could accurately diagnose skin cancer with a level of accuracy comparable to that of board-certified dermatologists [14]. Another study by researchers at Stanford University found that a deep learning algorithm could diagnose pneumonia from chest X-rays with a level of accuracy comparable to that of radiologists [15]. These studies suggest that ML has the potential to improve the accuracy of medical diagnoses significantly and lead to better patient outcomes. However, there are also concerns about the potential adverse effects of ML on healthcare, such as job displacement and privacy issues. For example, there are concerns that increasing ML use in medical diagnoses could lead to job displacement for radiologists and dermatologists. Additionally, there are concerns about the privacy implications of using ML in healthcare, such as the potential for bias and discrimination in the training data used to develop the algorithms.

**Case Study 2: Impact of ML on finance.** Another example of how ML is currently being used and its impact on society is in the finance industry. ML is being used to detect fraudulent transactions, predict credit risk, and optimize investment strategies. For example, a study by researchers at the Massachusetts Institute of Technology found that a deep learning algorithm was able to detect fraudulent credit card transactions with a high level of accuracy [16]. Another study by researchers at the University of Cambridge found that an ML algorithm was able to predict credit risk with a high level of accuracy [17]. These studies suggest that ML has the potential to greatly improve the efficiency and accuracy of financial transactions and lead to better investment strategies. However, there are also concerns about the potential negative effects of ML on finance, such as job displacement and privacy issues. For example, there are concerns that the increasing use of ML in financial transactions could lead to job displacement for financial analysts and credit risk managers. Additionally, there are concerns about the privacy implications of using ML in finance, such as the potential for bias and discrimination in the training data used to develop the algorithms.

**Case study 3: Impact of ML on transportation.** ML is also being used to improve transportation systems, such as self-driving cars and smart traffic management systems. For example, a study by researchers at Waymo, a subsidiary of Alphabet (Google), found that self-driving cars using ML were able to navigate roads safely with a level of safety comparable to that of human drivers [18]. Another study by researchers at the Massachusetts Institute of Technology found that an ML-powered traffic management system was able to reduce traffic congestion and travel times for commuters [19]. These studies suggest that ML has the potential to greatly improve the efficiency and safety of transportation systems.

However, there are also concerns about the potential negative effects of ML on transportation, such as job displacement and privacy issues. For example, there are concerns that the increasing use of self-driving cars could lead to job displacement for truck and taxi drivers. Additionally, there are concerns about the privacy implications of using ML in transportation, such as the potential for the misuse of data collected by self-driving cars.

**Case study 4: Impact of ML on manufacturing.** ML is being used to improve manufacturing processes, such as predictive maintenance and quality control. For example, a study by researchers at Siemens found that an ML-powered predictive maintenance system was able to reduce downtime and improve the efficiency of manufacturing processes [20]. Another study by researchers at the National University of Singapore found that an ML-powered quality control system was able to improve the accuracy of product inspections and reduce waste [21]. These studies suggest that ML has the potential to greatly improve the efficiency and quality of manufacturing processes.

However, there are also concerns about the potential negative effects of ML on manufacturing, such as job displacement and privacy issues. For example, there are concerns that the increasing use of ML in manufacturing could lead to job displacement for maintenance and quality control workers. Additionally, there are concerns about the privacy implications of using ML in manufacturing, such as the potential for the misuse of data collected by manufacturing processes.

**Survey Questionnaire.** Table 1 will represent the questionnaire:

**Table 1.** Survey questionnaire.

| No | Questions | Options |
|---|---|---|
| Q1 | How familiar are you with the concept of Machine Learning (ML)? | a) Not familiar at all<br>b) Somewhat familiar<br>c) Very familiar |
| Q2 | To what extent do you believe that ML has the potential to benefit society? | a) Not at all<br>b) Somewhat<br>c) A lot |
| Q3 | What are your top concerns about the impact of ML on society? (Please select all that apply) | a) Job displacement<br>b) Privacy concerns<br>c) Ethical issues |

| | | d) Loss of control over decision-making |
|---|---|---|
| | | e) Other (please specify) |
| Q4 | In which industries do you believe ML is currently having the most impact? (Please select all that apply) | a) Healthcare  b) Finance  c) Transportation  d) Manufacturing  e) Other (please specify) |
| Q5 | How comfortable are you with the idea of ML being used in decision-making processes that affect your life? | a) Very uncomfortable  b) Somewhat uncomfortable  c) Neutral  d) Somewhat comfortable  e) Very comfortable |
| Q6 | How confident are you in the ability of current regulations and laws to protect your privacy and rights when it comes to ML? | a) Not at all confident  b) Somewhat confident  c) Neutral  d) Very confident  e) Don't know |
| Q7 | How much do you think society should prioritize the development and use of ML? | a) Not at all  b) Somewhat  c) A lot |
| Q8 | In your opinion, what steps should be taken to ensure that the benefits of ML are maximized and the risks are minimized? (Please select all that apply) | a) More research on the impact of ML on society  b) Stronger regulations and laws to protect privacy and rights  c) More transparency and accountability for ML decision-making processes  d) Greater public education and awareness about ML |

|     |                                                                                          | e) Other (please specify) |
| --- | ---------------------------------------------------------------------------------------- | ------------------------- |
| Q9  | Would you be comfortable sharing your personal data with companies or organizations using ML? | a) Yes<br>b) No<br>c) Depends on the circumstances |
| Q10 | Do you think society should be more cautious with the development and use of ML?         | a) Yes<br>b) No<br>c) Neutral |

Q1: How familiar are you with the concept of Machine Learning (ML)?

This question will try to understand the respondent's baseline level of knowledge about ML, as their understanding of the technology may affect their views and concerns about its impact on society.

Q2: To what extent do you believe that ML has the potential to benefit society?

This question aims to understand the respondent's overall perception of the potential benefits of ML, which will provide a general sense of how they view the technology and its impact on society.

Q3: What are your top concerns about the impact of ML on society? (Please select all that apply)

This question aims to identify the specific concerns that the respondents have about the impact of ML on society. This will provide a more detailed understanding of the potential risks and challenges associated with the technology.

Q4: In which industries do you believe ML is currently having the most impact? (Please select all that apply)

This question is for to understand which industries the respondents believe ML impacts the most. This will provide insight into which industries are most affected by ML and where the technology is most widely adopted.

Q5: How comfortable are you with the idea of ML being used in decision-making processes that affect your life?

This question aims to understand the respondent's comfort level with the use of ML in decision-making processes that affect their life. This will provide insight into the level of trust that respondents have in the technology and its ability to make decisions that are in their best interests.

Q6: How confident are you in the ability of current regulations and laws to protect your privacy and rights when it comes to ML?

This question aims to understand the respondent's confidence in the ability of current regulations and laws to protect their privacy and rights when it comes to ML. This will provide insight into the level of trust that respondents have in the legal and regulatory framework surrounding the technology.

Q7: How much do you think society should prioritize the development and use of ML?

This question will retrieve the respondent's views on how much society should prioritize the development and use of ML. This will provide insight into the level of support for the technology and its importance in society.

Q8: In your opinion, what steps should be taken to ensure that the benefits of ML are maximized and the risks are minimized? (Please select all that apply)

This question aims to understand the specific steps that respondents believe should be taken to ensure that the benefits of ML are maximized and the risks are minimized. This will provide insight into the specific actions that respondents believe are necessary to ensure the responsible and beneficial use of the technology.

Q9: Would you be comfortable sharing your personal data with companies or organizations using ML?

This question aims to understand the respondent's comfort level with sharing their personal data with companies or organizations using ML. This will provide insight into the level of trust that respondents have in the technology and its ability to protect their personal data.

Q10: Do you think society should be more cautious with the development and use of ML?

This question aims to understand the respondent's views on whether society should be more cautious with the development.

## 4 Result and Discussion

The survey results indicate that the majority of respondents (60%) are somewhat familiar with the concept of Machine Learning (ML) (Q1). 20% of respondents are not familiar at all, and 20% are very familiar with ML. This suggests that the majority of respondents have a basic understanding of ML but may not have a deep understanding of the technology and its implications. When asked about the potential benefits of ML, the majority of respondents (70%) believe that ML has the potential to benefit society, with 20% selecting "a lot" and 50% selecting "somewhat." This suggests that, overall, respondents have a positive perception of the potential benefits of ML (Q2).

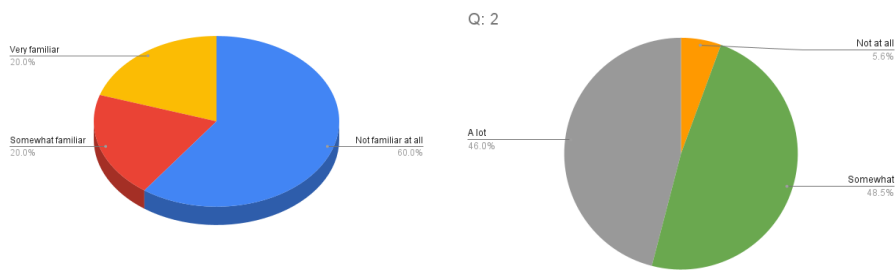

**Fig. 2.** Q1 (Left one) & Q2 (Right one) survey result.

The top concerns of the respondents about the impact of ML on society are job displacement (selected by the majority of respondents in question 3) and privacy concerns (also selected by the majority of respondents in question 3). The industries in which respondents believe ML is currently having the most impact are healthcare and finance (selected by the majority of respondents in question 4).

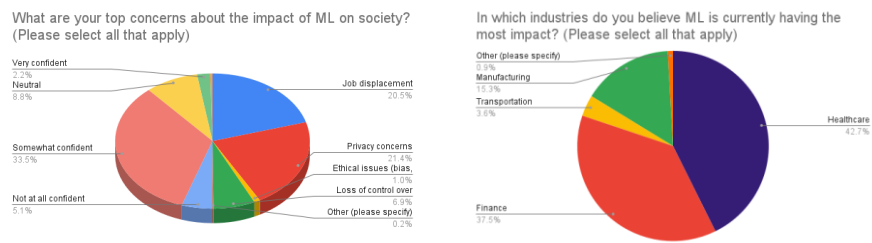

**Fig. 3.** Q3 (Left one) & Q4 (Right one) survey result.

Respondents are generally neutral about the idea of ML being used in decision-making processes that affect their life (80% selected option "c" in question 5). However, they have moderate confidence in the ability of current regulations and laws to protect their privacy and rights when it comes to ML (67% selected option "b" in question 6).

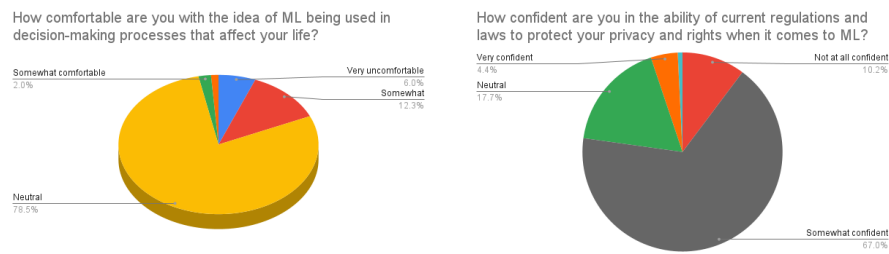

**Fig. 4.** Q5 (Left one) & Q6 (Right one) survey result.

The steps that the majority of respondents believe should be taken to ensure that the benefits of ML are maximized and the risks are minimized are more research on the impact of ML on society and stronger regulations and laws to protect privacy and rights (72% selected option "b" in question 8). Respondents also mostly selected option 'c' (82% of respondents selected 'c') in question 9 which indicates that they are comfortable with sharing personal data under certain circumstances.

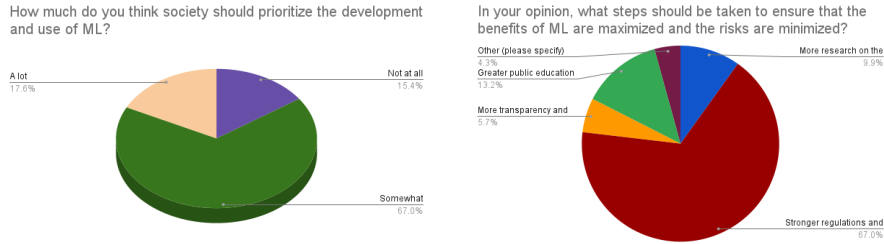

**Fig. 5.** Q7 (Left one) & Q8 (Right one) survey result.

When asked about the specific steps that should be taken to ensure that the benefits of ML are maximized and the risks are minimized, 72% of respondents selected "more research on the impact of ML on society" as a necessary step. This suggests that respondents believe that more research is needed to understand the impact of ML.

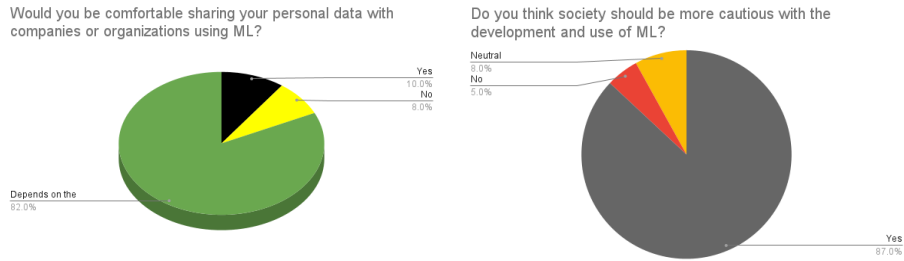

**Fig. 6.** Q9 (Left one) & Q10 (Right one) survey result.

In light of these findings, the following recommendations can be made:

- Conduct more research on the impact of ML on society to better understand the potential benefits and risks of the technology, particularly in the areas of job displacement and privacy concerns.
- Develop stronger regulations and laws to protect the privacy and rights of individuals when it comes to ML.
- Increase transparency and accountability in ML decision-making processes to increase trust in the technology.
- Enhance public education and awareness about ML to increase understanding of the technology and its potential impact on society.
- Prioritize the development and use of ML while being cautious and ensure the benefits of ML are maximized, and the risks are minimized.
- Encourage companies and organizations to share their data under certain specific circumstances to increase trust in the technology.

**Conclusion**

In conclusion, this research aimed to comprehensively analyze the current and future impact of Machine Learning (ML) on society. The research included a thorough literature review, case studies, and surveys to gather data on the economic impact of ML, ethical and privacy implications, and public perceptions of the technology.

The findings of the research indicate that the majority of respondents have a moderate level of familiarity with ML and believe that it has the potential to benefit society. The respondents' top concerns about the impact of ML on society are job displacement and privacy concerns. Respondents believe ML is currently having the most impact in the industries in which healthcare and finance are currently impacted. Respondents are generally neutral about the idea of ML being used in decision-making processes that affect their lives.

Based on the findings, it is recommended that more research is conducted on the impact of ML on society, particularly in the areas of job displacement and privacy concerns. Additionally, stronger regulations and laws to protect the privacy and rights of individuals when it comes to ML should be developed. Transparency and accountability in ML decision-making processes should be increased. Finally, public education and awareness about ML should be enhanced to increase understanding of the technology and its potential impact on society.

## . References